\documentclass[twocolumn]{aastex62}

\turnoffeditone

\usepackage{graphicx}                    
\usepackage{color}                       
\usepackage{breakurl}                    

\usepackage{amsmath}

\graphicspath{{./}{Figures/}}

\shorttitle{3D density structure of a coronal streamer}
\shortauthors{B. Decraemer et al.}

\begin{document}

\title{Three-dimensional Density Structure of a Solar Coronal Streamer Observed by SOHO/LASCO and STEREO/COR2 in Quadrature}

\correspondingauthor{Bieke Decraemer}
\email{bieke.decraemer@oma.be}

\author[0000-0002-1335-7768]{Bieke Decraemer}
\affiliation{Solar-Terrestrial Centre of Excellence - SIDC, Royal Observatory of Belgium, Ringlaan 3, 1180 Brussels, Belgium}
\affiliation{Centre for mathematical Plasma Astrophysics - CmPA, Mathematics Department, KU Leuven, Celestijnenlaan 200B, 3001 Leuven, Belgium}

\author[0000-0002-2542-9810]{Andrei N. Zhukov}
\affiliation{Solar-Terrestrial Centre of Excellence - SIDC, Royal Observatory of Belgium, Ringlaan 3, 1180 Brussels, Belgium}
\affiliation{Skobeltsyn Institute of Nuclear Physics, Moscow State University, 119992 Moscow, Russia}

\author[0000-0001-9628-4113]{Tom Van Doorsselaere}
\affiliation{Centre for mathematical Plasma Astrophysics - CmPA, Mathematics Department, KU Leuven, Celestijnenlaan 200B, 3001 Leuven, Belgium}

\begin{abstract}
Helmet streamers are a prominent manifestation of magnetic structures with current sheets in the solar corona. These large-scale structures are regions with high plasma density, overlying active regions and filament channels. We investigate the three-dimensional (3D) structure of a coronal streamer, observed simultaneously by white-light coronagraphs  from two vantage points near quadrature (SOHO/LASCO and STEREO/COR2). We design a forward model based on plausible assumptions about the 3D streamer structure taken from physical models (a plasma slab centered around a current sheet). The streamer stalk is approximated by a plasma slab, with electron density that is characterized by three separate functions describing the radial, transverse and face-on profiles respectively. For the first time, we simultaneously fit the observational data from SOHO and STEREO using a multivariate minimization algorithm. The streamer plasma sheet contains a number of brighter and darker ray-like structures with the density contrast up to about a factor 3 between them. The densities derived using polarized and unpolarized data are similar. We demonstrate that our model corresponds well to the observations.
\end{abstract}




%
\section{Introduction}\label{s:introduction} 
Observations of the solar corona during total eclipses reveal a striking ray-like view of the middle and outer solar corona  \citep[see e.g.][]{loucif_solar_1989}. Many rays present a characteristic shape: a bulge or helmet-like feature close to the Sun, narrowing to a very thin stalk further out \citep{koutchmy_coronal_1992}. These very bright narrow rays, called helmet streamers, trace out the global magnetic field configuration of the Sun. The brightness inferred from the solar corona in white-light observations is mainly the result of Thomson scattering of photospheric light on free electrons in the corona \edit1{\added{\citep[e.g.][]{inhester2016}}}. The bright streamers thus indicate where regions of high plasma density can be found. Typically, helmet streamers are large-scale structures found overlying active regions and filament channels, which often lie above a photospheric neutral line that often coincides with the heliospheric current sheet \citep[see e.g.][]{newkirk_structure_1967, koutchmy_three_1971, zhukov_origin_2008}.

Observations of streamers have much improved since the  1970s with the launch of space-borne coronagraphs. A very large improvement came with the Large Angle Spectroscopic Coronagraph aboard the \textit{Solar and Heliospheric Observatory} \citep[LASCO aboard \textit{SOHO}; see][]{brueckner_large_1995}. A number of models for the electron density distributions were developed based on the high-resolution observations of the total brightness of the white-light corona from the LASCO coronagraphs \citep{koomen_shape_1998, wang_dynamical_2002, gibson_three-dimensional_2003, saez_3-dimensional_2005, saez_three-dimensional_2007, morgan_empirical_2007}. Typically, only one vantage point was used in combination with the solar rotation. This implies that only slowly-varying features can be correctly captured. Consequently, the focus of modeling has mostly been on the large-scale corona which has more long-lived features. However, \citet{eselevich_investigation_1999} and \citet{thernisien_electron_2006} have shown that the streamer belt clearly has many fine ray-like structures. Moreover, even large structures like helmet streamers can be disrupted by coronal mass ejections (CMEs) and thus only exist for a short time.  

Most recently, we have the addition of the twin spacecraft (A and B) of the \textit{Solar Terrestrial Relations Observatory} (\textit{STEREO}) mission \citep{kaiser_stereo_2008}. There are two coronagraphs, COR1 and COR2, aboard each spacecraft. They are part of the Sun Earth Connection Coronal and Heliospheric Investigation (SECCHI) instrument package \citep{howard_sun_2008}.  The combination of coronagraph observations from different vantage points gives the possibility of viewing the coronal structure from different angles, since the STEREO~A (B) spacecraft moves in an orbit around the Sun ahead (behind) the Earth, respectively, while SOHO remains in the L1 point between the Earth and the Sun. The different viewing angles create ideal opportunities for three-dimensional (3D) reconstructions of coronal structures.  


The 3D reconstruction of coronal features typically is a two-part problem: on the one hand, one has the geometric structure to be inferred with a correct estimation of projection effects, on the other hand, one needs to perform the inversion of the perceived brightness to electron density. The problem of reconstruction can be approached with different methods, each with their own assumptions and limitations. Tomography is a valuable inverse method to make reconstructions of the electron density in the large-scale corona and the streamer belt, though it typically needs continuous observations for multiple days \citep[see e.g.][]{vasquez_validation_2008, frazin_three-dimensional_2010, vasquez_whi_2011, aschwanden_solar_2011}. More recently, \citet{morgan2015} and \citet{morgan2019} have developed a novel tomography method to reconstruct qualitative global coronal density maps at a specific height in the inner corona. By identifying the same feature in at least two different views, triangulation methods or tie-point reconstructions can provide the location of coronal features in 3D. This technique has already been demonstrated for coronal loops \citep{aschwanden_first_2008-1} and CMEs \citep{mierla_3d_2009}. By incorporating stereoscopic images from different wavelengths, the electron density was derived by \citet{aschwanden_first_2008}. \citet{minnaert_continuous_1930} and \citet{van_de_hulst_electron_1950} laid out the foundations for the forward modeling technique to estimate the electron density radial profile in coronal streamers and coronal holes. Usually based on physical assumptions, a specific model is chosen to represent the 3D geometry of the feature. This reduces the inverse problem to the determination of the free parameters in the model. \citet{gibson_three-dimensional_2003}, \citet{thernisien_modeling_2006} and \citet{thernisien_forward_2009} have expanded this technique to make three-dimensional models of the solar corona, to work with non-spherically symmetric structures, and to use it for dynamical events such as CMEs.

In this paper, we investigate the 3D structure of a coronal streamer, observed by the white-light coronagraphs SOHO/LASCO and STEREO/COR2 from two vantage points located close to the quadrature. In Section~\ref{s:data}, we describe the different observations used to determine the location of the streamer and how we pre-processed the data. We present our forward modeling method of fitting a simple slab model to the streamer with a multivariate minimization technique in Section~\ref{s:fitting}. In Section~\ref{s:discussion}, we discuss and compare our results with each other and the observations, and with other electron density models. We present our conclusions in Section~\ref{s:conclusions}.  

%
%
\begin{figure*}
    \centering
    \plottwo{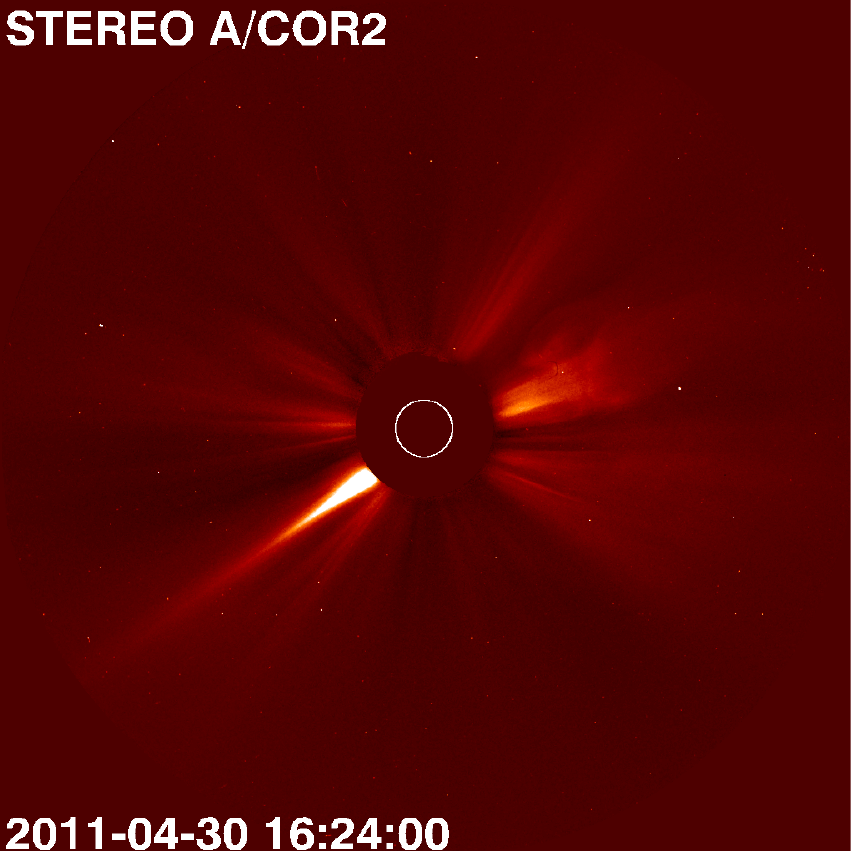}{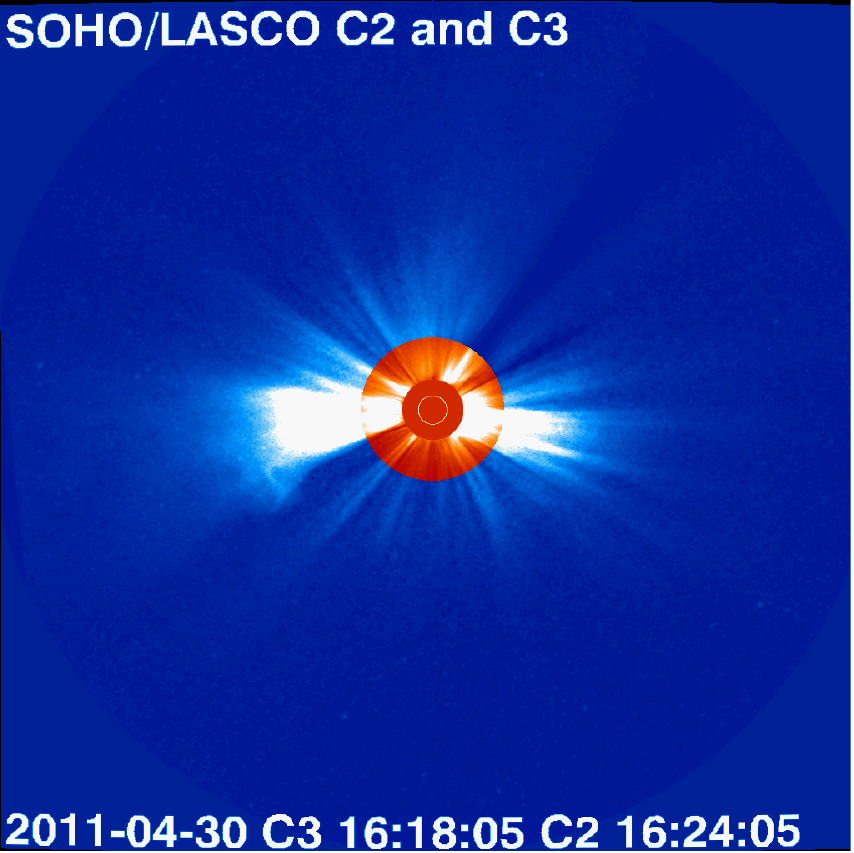}
    \caption{STEREO~A/COR2 and SOHO/LASCO images showing respectively the edge-on and face-on views of the streamer on April 30, 2011. The streamer is located above the south-east limb in the COR2 image and above the south limb in the LASCO image. The right panel is a composite image of the C2 and C3 coronagraph field of view (FOV) and goes out to $30\ R_{\sun}$. The FOV of COR2 is $15\ R_{\sun}$ (left panel).}
    \label{fig:views}
\end{figure*}
\section{Data}\label{s:data}
\begin{figure} 
	\centering
	\plotone{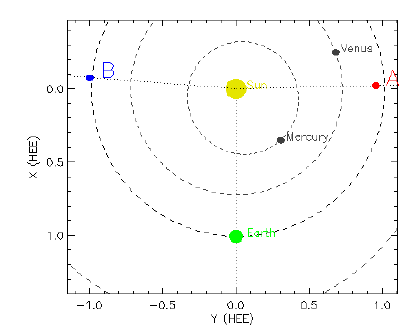}
	\caption{Positions of STEREO~A and B, and Earth for 2011-04-30 16:24 UT (courtesy STEREO Science Center, \url{https://stereo-ssc.nascom.nasa.gov/cgi-bin/make_where_gif}).}		\label{fig:stereo_orbit}
\end{figure}
On April 30, 2011 a clear streamer structure was observed above the south-east limb by the COR2 coronagraph aboard STEREO~A  (see Figure \ref{fig:views}, left panel). At the time of observation the STEREO~A spacecraft and Earth were separated by an angle of 91.415\degr{}, as can be seen in Figure \ref{fig:stereo_orbit}. Since SOHO is in the vicinity of the Sun-Earth Lagrange L1 point, the separation angle between STEREO~A and SOHO was very close to 90\degr{} as well. Since STEREO B at this time was almost exactly opposite of STEREO~A with respect to the Sun, the view from COR2 B is redundant and not used in this paper. The choice of COR2 A was made, since COR2 A has a better signal-to-noise ratio than COR2 B. The streamer structure is therefore observed by two spacecraft carrying coronagraphs nearly in quadrature, which makes this situation very well suited for three-dimensional reconstructions.
\begin{figure*}
    \centering
    \plottwo{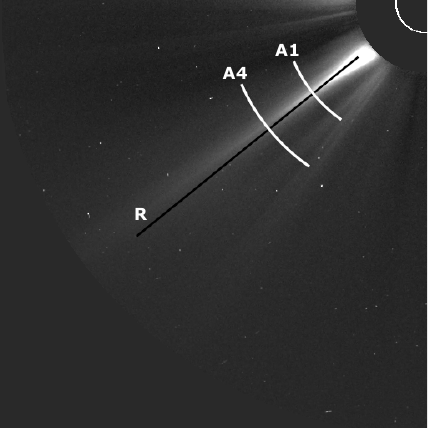}{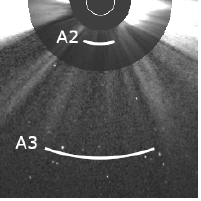}
    \caption{STEREO~A/COR2 (left) and SOHO/LASCO C2+C3 (right) zoomed images showing respectively the edge-on and face-on views of the streamer on April 30, 2011. R, A1, A2, A3, and A4 are the radial and arc-shaped lines along which brightness profiles are extracted for the fitting procedure.}
    \label{fig:extractions}
\end{figure*}
The view from COR2 shows a narrow streamer stalk around a latitude of -39\degr{}, located in the south-east quadrant. It presents the typical image we have for coronal streamers and which we will call henceforth the \textit{edge-on} view (Figure~\ref{fig:extractions}, left panel). With the LASCO C2 and C3 coronagraphs, we have the view rotated by approximately 90\degr{}, which we will refer to as the \textit{face-on} view. In the face-on view in LASCO, the streamer has a fan-like structure with radially extended regions of higher and lower brightness, see the right panels in Figures~\ref{fig:views} and~\ref{fig:extractions}. The density along the azimuthal direction is clearly not uniform.

We use both unpolarized and polarized images to reconstruct the coronal electron density. The first set of images used in this study was taken at 16:24:00 UT by COR2 using so-called ``double'' (i.e. total brightness) exposure \citep[see][]{howard_sun_2008}, at 16:23:04 UT by LASCO C2 and at 16:17:04 UT by LASCO C3. These are total brightness images. We also use a set of polarized brightness (\textit{pB}) images, namely the \textit{pB} image sequence from COR2 at 16:08:15 UT, 16:08:45 UT and 16:09:15 UT, and the \textit{pB} image sequence for LASCO C2 at 14:54:08 UT, 14:57:58 UT and 15:01:48 UT.

\subsection{Background removal}\label{s:calibrating}
As a first step, the data needs to be correctly pre-processed and calibrated to separate the K corona from the F corona and straylight. We need calibrated data in units of mean solar brightness (MSB), since our model will be fitted to intensity values, which correspond in turn to the line-of-sight integration of electron density values.

We use the total brightness data, prepared using the standard procedures \verb|reduce_level_1.pro| and \verb|secchi_prep.pro| in SolarSoft\footnote{\url{http://www.lmsal.com/solarsoft/}} and the appropriate subroutines for the \textit{pB} images. The pre-processing procedures most importantly subtract the offset bias, correct for exposure time, and calibrate the data to obtain physical units (MSB). For the total brightness images, we also need to remove the background that contains the straylight and the F-corona. To remove the background from the total brightness images, we use a monthly minimum image. We follow techniques implemented earlier for SOHO/LASCO and STEREO/COR1 \citep{morrill_calibration_2006, thompson_background_2010}. For each day in a 28-day period around our chosen date, we select at random 40 images from the available image sets. For the 28 sets of 40 images, the images are processed, as described above. However, they are not yet calibrated to physical units. Then we find the median brightness in each pixel over the 40 images. This gives us 28 daily median images. Next, we take the minimum brightness in each pixel from the 28 median images. Finally, this monthly minimum background image is subtracted and the resulting image is calibrated to units of MSB. The background image will also contain the fraction of the K corona that remains constant during the full rotation. Subtracting this background from our total brightness images gives us a reasonable approximation of the dynamic K corona intensity, but will inevitably underestimate the total K corona brightness, and thus the electron density derived from our method.

Electron density inversion methods have also been widely used with polarized brightness (\textit{pB}) images, since the F corona is unpolarized below approximately 5~$R_{\sun}$ \citep{quemerais_two-dimensional_2002}. The K corona can thus be directly inferred from \textit{pB} images in the regions below 5~$R_{\sun}$. Above this height the polarization of the F corona cannot be neglected \citep{koutchmy_f-corona_1985, mann_solar_1992}, so using  a \textit{pB}-based method overestimates the electron density in this region. We present the results using a full set of \textit{pB} images in Section~\ref{s:pbparameters} below. In Section~\ref{s:comparisontwomodels}, we compare the results of the two input methods. 

An important argument in favor of using the total brightness data is that the cadence of total brightness images is often significantly higher than for \textit{pB} images for the white-light coronagraphs used here. For example, LASCO C2 takes only a few polarized image sequences per day, while the cadence of unpolarized images is typically 12--20 minutes. If our model would be used as a direct density input to study more dynamical events, we need the highest cadence available to capture as many features as possible. Examples of transient events that could be studied are streamer blow-out CMEs \citep[see e.g.][]{vourlidas_streamer-blowout_2018} and streamer waves. Streamer waves are propagating waves along a coronal streamer, typically caused by a CME hitting and displacing the streamer \citep{chen_streamer_2010, feng_streamer_2011}.

\subsection{Position of the streamer}\label{s:position}
\begin{figure*} 
	\centering
	\plotone{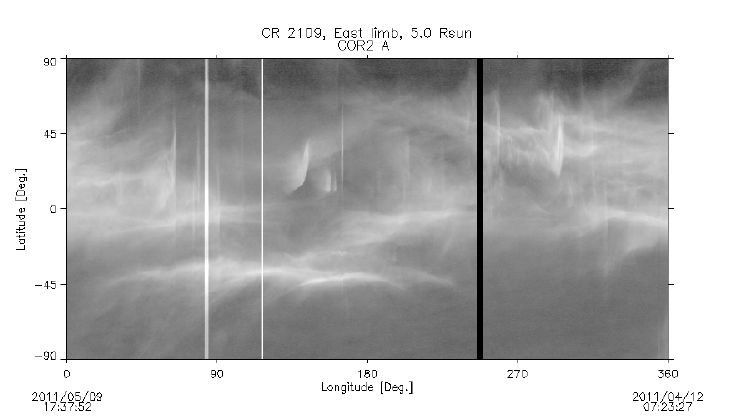}
	\caption{Synoptic map for CR 2109 made at 5.0~$R_{\sun}$ above the east limb from COR2 A data images (from \url{https://secchi.nrl.navy.mil/synomaps/}). Black and white vertical lines are due to missing or bad data. The COR2 image in Figure~\ref{fig:views} corresponds to the Carrington longitude of 118\degr{} at the east limb.}\label{fig:synopticmap}
\end{figure*}
In order to determine the position and geometry of the streamer in 3D space, we use the common method of comparison of coronagraphic synoptic maps with the extrapolated coronal magnetic field \citep[e.g.][]{wang_evolution_2000, saez_3-dimensional_2005, saez_three-dimensional_2007, zhukov_origin_2008}. In the COR2 A synoptic map, shown in Figure~\ref{fig:synopticmap} for Carrington rotation 2109 (CR 2109), the streamer structure spans about 40\degr{} in longitude at a latitude of around -40\degr{}, between Carrington longitudes 150\degr{} and 110\degr{}. The characteristic curvature at the edges of the streamer track towards the pole is due to the projection effect appearing during the rotation of the Sun \citep{wang_evolution_2000}. To get a more accurate latitude of the narrow streamer, we extract a circular profile from the edge-on view at 5.0 $R_{\sun}$ (shown as curve A1 in Figure~\ref{fig:extractions}) and locate the peak of brightness in this profile. This gives us the location of the central streamer axis at -39\degr{} latitude at the east limb, or a position angle (P.A.) of 129\degr{}, where the solar north coincides with a P.A. of 0\degr{}. The radial line R on Figure~\ref{fig:extractions} shows the streamer axis according to this method. 

\begin{figure*} 
	\centering
	\plotone{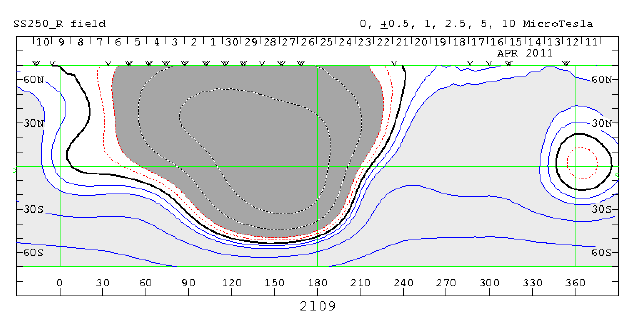}
	\caption{WSO source surface field map from the PFSS radial model at $2.5\ R_{\sun}$ for CR 2109 (from April 12 to May 9, 2011). The part of the neutral line in the southern hemisphere at the latitude around 50\degr{} and between longitudes 150\degr{} and 110\degr{} corresponds to the studied streamer. (from \url{http://wso.stanford.edu/synsourcel.html})} 		\label{fig:WSOmap}
\end{figure*}

Figure~\ref{fig:WSOmap} shows the potential field source surface (PFSS) map of the extrapolated coronal magnetic field at 2.5~$R_{\sun}$ from the Wilcox Solar Observatory (WSO) photospheric magnetogram. In the southern hemisphere, between around 110\degr{} and 150\degr{} longitude, the neutral line remains parallel to the equator at a constant latitude of about 50\degr{}. Since STEREO and SOHO are separated by about 90\degr{}, we can infer from the source surface map that the narrow streamer seen in the COR2 view corresponds to this flat part of the source surface neutral line seen edge-on. This indicates that the streamer is centered around a current sheet in the corona, and thus is a true helmet streamer as opposed to a pseudo-streamer \citep{2007ApJ...658.1340W}. The latitude of the part of the neutral line parallel to the equator (around 50\degr{}) does not completely agree with the latitude we get from the COR2 synoptic map and the brightness peak in the circular profile around 40\degr{}. This difference of only 10\degr{} is probably due to the inaccuracy of the PFSS model in the determination of the location of the neutral line \citep[see e.g.][]{zhukov_origin_2008}. We will come back to this issue later on in this paper. 

We thus approximate the 3D configuration of our streamer by a slab-like structure, having an angular extent of 40\degr{}, essentially in Carrington longitude, in the face-on view centered around the central axis in the southern hemisphere, at a position angle of 129\degr{} in the edge-on view.

%
%
\section{Fitting the observations with the model}\label{s:fitting}
Using observations from several vantage points, a realistic reconstruction of the 3D configuration can be made for coronal streamers. We develop a forward model based on plausible assumptions about the 3D streamer structure derived in Section~\ref{s:position}. Our model is based upon the model presented by \citet{thernisien_electron_2006}. It has already been determined that a slab model is a reasonable approximation of the 3D structure of a streamer \citep{guhathakurta_large-scale_1996,vibert_streamer_1997,thernisien_electron_2006}. The major improvement in this work is that we for the first time use truly simultaneous views from different vantage points. Previous works had to use the solar rotation to obtain the face-on and edge-on view, or only compare simulated and observed synoptic maps.  

The narrow streamer will be modeled as a plasma slab, located at the position we derived above. A great advantage of this kind of model is that we can work with an orthogonal set of parameters. That is, we can separate the radial, azimuthal, and latitudinal model parameters and fit them to the observations independently. This makes the electron density derivation much more straightforward and computationally less expensive than e.g. tomographic inversion methods \citep[e.g.][]{vasquez_validation_2008}. 

\subsection{The slab model}\label{s:model}
The slab model considered for our streamer is built up from the multiplication of three functions. They describe respectively the radial, transverse and face-on profiles for the electron density $n_{\mathrm{e}}$. It can be described by the following equation \citep{thernisien_electron_2006}:
\begin{equation}
n_{\mathrm{e}}(r, \alpha, \theta) = n_{\mathrm{e, radial}}(r)n_{\mathrm{e, shape}}(r, \theta)n_{\mathrm{e, face}}(\alpha),
\end{equation}
where $r$ is the radial distance (in solar radii), $\alpha$ is the azimuthal angle in the plane of the slab and $\theta$ is the latitudinal angle between the line defined by a point in the corona and the Sun center to the plane of the slab, see Figure~4 in \citet{thernisien_electron_2006}. The three $n_{\mathrm{e}}$ functions (radial, shape, and face) from \citet{thernisien_electron_2006}, in the order they appear in the fitting process, are here repeated for convenience.

The shape function $n_{\mathrm{e, shape}}$ is specified by
\begin{align}\label{eq:shape}
&n_{\mathrm{e, shape}} (r, \theta) = \nonumber \\
&\begin{cases}
\exp\left( -\dfrac{\theta^2}{\theta_1(r)}\right) , &\left\vert\theta\right\vert < \dfrac{\theta_1(r)}{2\theta_2(r)},\\
\exp\left( -\left\vert\dfrac{\theta}{\theta_2(r)}\right\vert + \dfrac{\theta_1(r)}{4 \theta_2(r)^2} \right) , &\left\vert\theta\right\vert \geq \dfrac{\theta_1(r)}{2\theta_2(r)}.
\end{cases}
\end{align}
This function represents how the thickness of the slab varies with the radial distance, i.e., how the shape of the narrow streamer, seen edge-on, changes. The 10 parameters to be determined are identified within the $\theta_1$ and $\theta_2$ functions, which are given by
\begin{subequations}
\begin{align}
\theta_1(r) &= \sum_{i=0}^4 b_i r^{-i},\\
\theta_2(r) &= \sum_{i=0}^4 c_i r^{-i},
\end{align}
\end{subequations}
where we have to find $b_i$ and $c_i$ in the optimization.

The radial dependence of the electron density is given by the $n_{\mathrm{e, radial}}$ function. It is given by the following equation:
\begin{equation}
n_{\mathrm{e, radial}}(r) = \sum_{i=1}^4 a_i r^{-i},
\end{equation}
and has four coefficients that need to be fitted to the observations, in contrast to the five coefficients for the polynomials that we have to find for both the $\theta_1$ and $\theta_2$ function.

Finally, the face-on modulation is expressed by simply inserting the normalized circular profile at 3 $R_{\sun}$, taken from the face-on brightness along the angular extent of the slab. This is different from the modulation implemented in \citet{thernisien_electron_2006}, where a linearization between two brightness profiles along the face-on view was used. The $n_{\mathrm{e,face}}$ function will enhance or reduce the electron density in the plane of the slab (along the current sheet), in order to reproduce the structure with a number of radially extended rays that we observe in the face-on view (see Figure~\ref{fig:views}, right panel). With this approach, we consider all the ray-like structures that are observed to be in the plasma slab and assume that they are radial features. 


\subsection{Fitting the total brightness images}\label{s:parameters}
\begin{figure}
    \centering
    \epsscale{1.1}
    \plotone{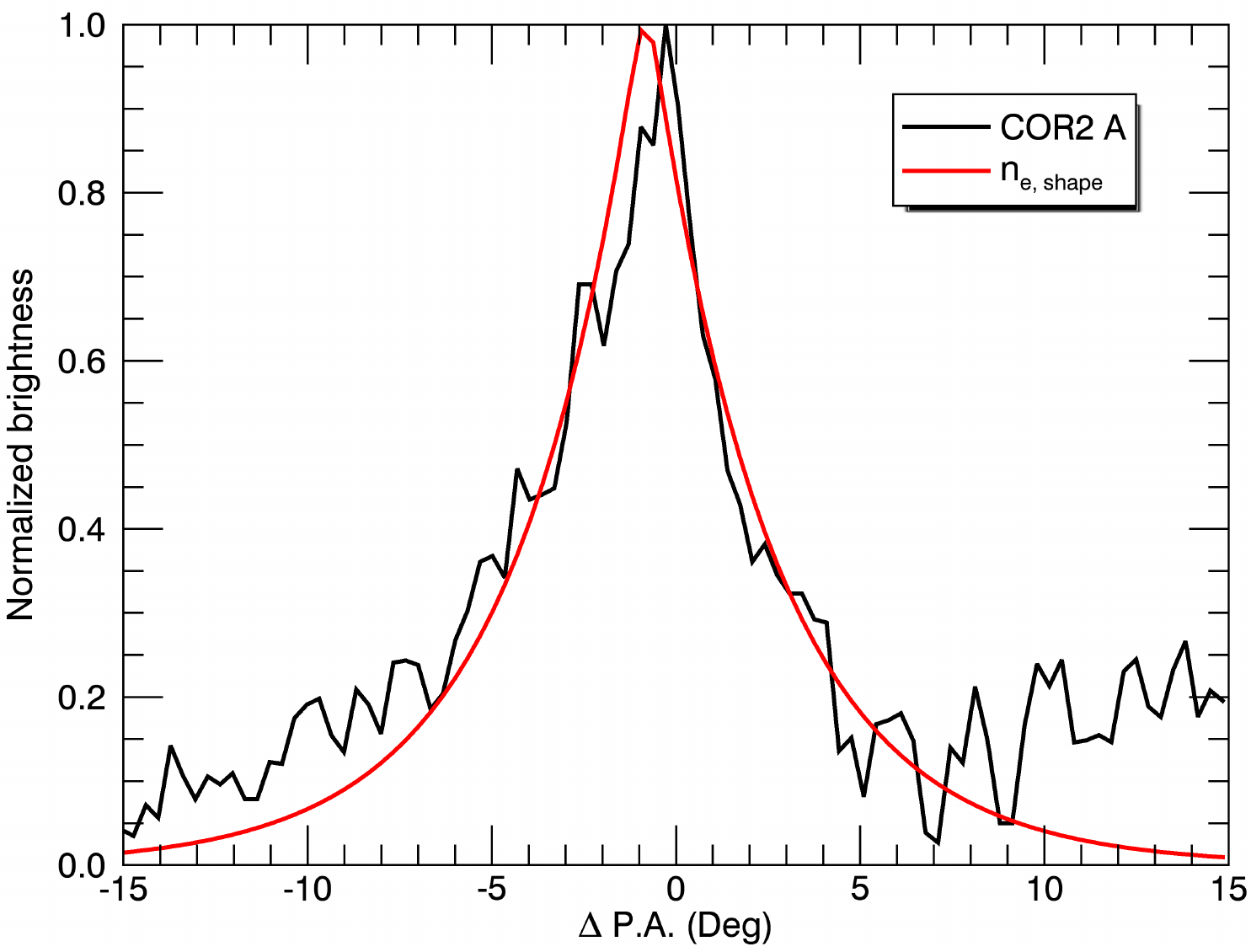}
    \caption{Normalized profiles of the brightness of the streamer viewed edge-on at 5 $R_{\sun}$ in the STEREO~A/COR2 data (along the line A1 in Figure~\ref{fig:extractions}) and in the model. The black line represents the observed brightness. The red line is the $n_{\mathrm{e,shape}}$ function, calculated using the optimized parameters. In this and other figures, the position angle is measured with respect to the position angle of 129\degr{}, which was determined to be the position of the central axis of the streamer slab in Section~\ref{s:position}.}
    \label{fig:5Rsun}
\end{figure}

We start with the fitting process by the slab model using the set of total brightness images, pre-processed as outlined in Section~\ref{s:calibrating}.
First, the shape function is fitted to arc-shaped profiles taken from the edge-on observations. We have used seventeen arc-shaped profiles taken at different heights in the COR2 field of view (FOV), ranging from 3 to 10~$R_{\sun}$ with intervals of half a solar radius and at 11 and 12~$R_{\sun}$. Curve A1 shown in Figure~\ref{fig:extractions} is an example of such an arc-shaped profile at 5~$R_{\sun}$. The bright structure at the right of the streamer that can be seen in the left panel of Figure~\ref{fig:views} around the latitude of 50\degr{} must be excluded from the data in the fitting process, since it does not contribute to the streamer profile. The profiles are normalized so that they can be fitted directly to the $n_{\mathrm{e, shape}}$ function. By doing the fit directly with the electron density function, we neglect any projection effects in this viewing direction. This approach is reasonable since the streamer slab is oriented orthogonal to the plane-of-the-sky. The coefficients $b_i$ and $c_i$, corresponding to the functions $\theta_1$ and $\theta_2$, are determined with an MPFIT minimization technique on a $\chi^2$ criterion \citep{markwardt_non-linear_2009}. The profiles at all heights are fitted simultaneously. An example of the fit at 5~$R_{\sun}$ can be seen in Figure \ref{fig:5Rsun}. The black line shows the normalized profile extracted from the COR2 A observations along the line A1 in Figure~\ref{fig:extractions}. The red line represents the $n_{\mathrm{e,shape}}$ function calculated on the base of the optimized parameters. The shape function can be directly compared with normalized observations since it is dimensionless, as can be seen in Equation~\ref{eq:shape}. The optimized shape function matches the observations reasonably well. 

\begin{figure*}
    \centering
    \plottwo{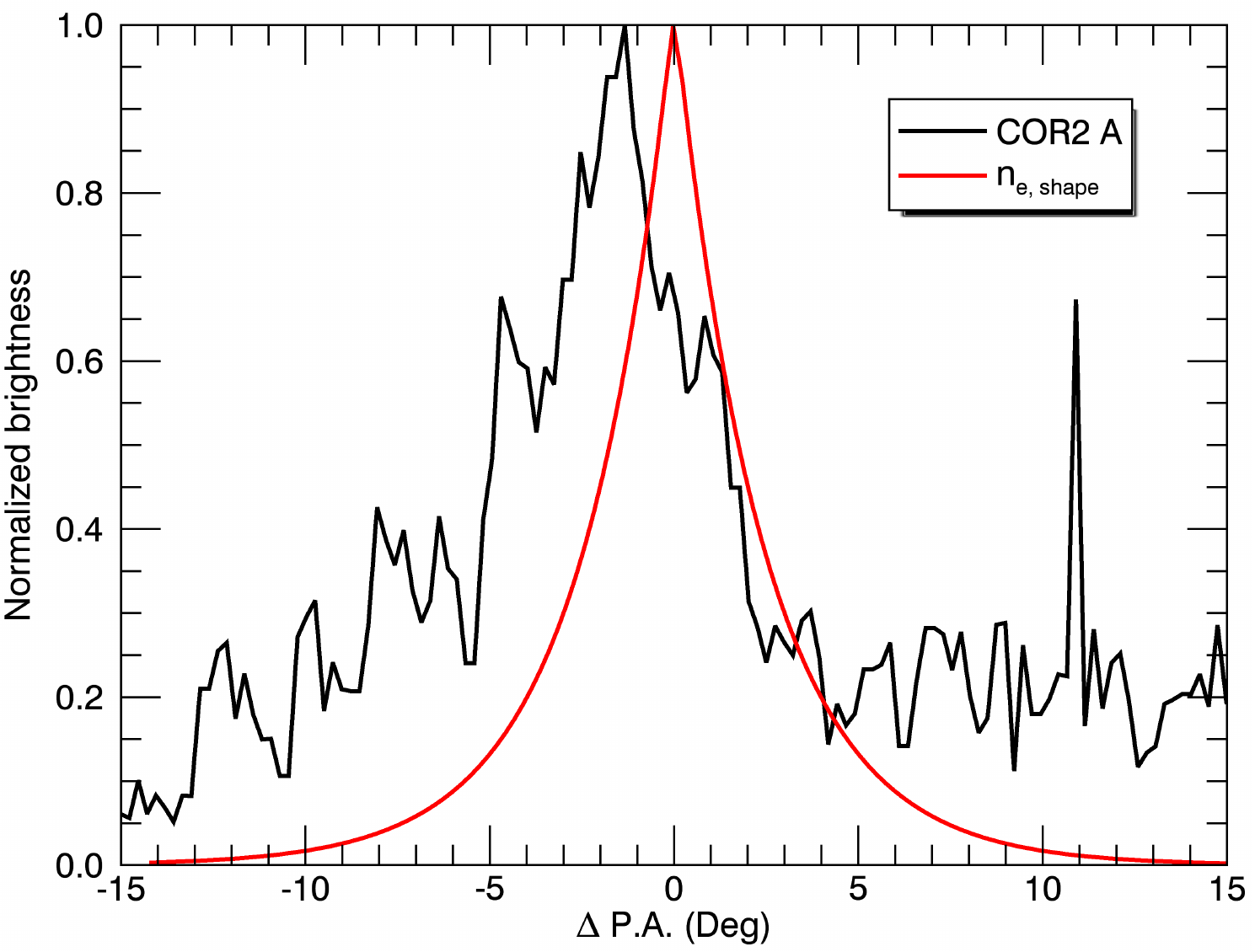}{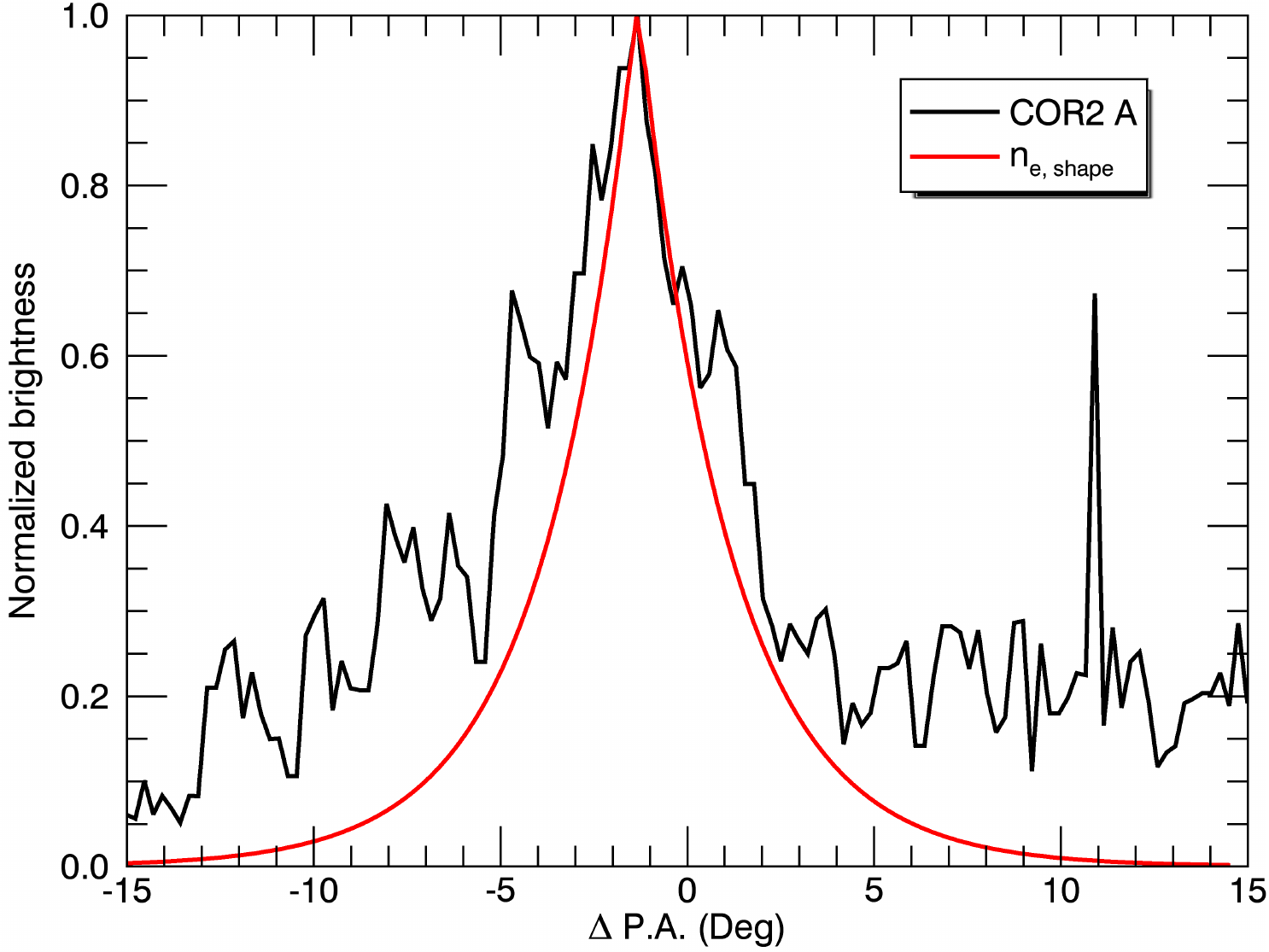}
    \caption{Normalized profiles of the brightness of the streamer viewed edge-on at 7 $R_{\sun}$ in the STEREO~A/COR2 data (along the line A4 in Figure~\ref{fig:extractions}) and in the model. The black line represents the observed brightness. The red line is the $n_{\mathrm{e,shape}}$ function, calculated using the optimized parameters. On the left, $n_{\mathrm{e,shape}}$ is centered at the streamer axis as determined at 5 $R_{\sun}$ (see section \ref{s:position}), on the right the center of $n_{\mathrm{e,shape}}$ is shifted by -1.34\degr{} in the P.A. to match the peak in brightness at 7 $R_{\sun}$.}
    \label{fig:7Rsun}
\end{figure*}

In Figure~\ref{fig:7Rsun} we show the shape function and observed coronal brightness profile at 7~$R_{\sun}$ along the line A4 in Figure~\ref{fig:extractions}. In the definition of the shape function, it is assumed that the axis of the streamer in the edge-on view is radial, due to the symmetry in the $\theta$ coordinate. We see however in Figure~\ref{fig:7Rsun} that the center of the streamer at 7~$R_{\sun}$ is not in the same location as the one that we derived through fitting the peak of brightness at 5~$R_{\sun}$. The shape of the streamer reasonably agrees with the observations if we shift the $n_{\mathrm{e, shape}}$ function by -1.34\degr{} to match the center with the peak in brightness of the observations (Figure~\ref{fig:7Rsun}, right panel). To compensate for the offset of the streamer axis, we determine the peak of brightness at each height chosen for the fitting of the shape function. Then we take the average of the peak location as the fixed location of the streamer axis, around which we assume the streamer to be symmetrical. The new P.A. of the streamer axis for the fitting procedure is 128.17\degr{}, as can be seen by the slight offset of the peak of $n_{\mathrm{e, shape}}$ to the center of the plot in Figure~\ref{fig:5Rsun}. 

Since the brightness peak at larger heights is located closer to the equator, the streamer appears to be bent towards the equator. This indicates that we are dealing with a non-radial streamer, which our model unfortunately can not reproduce. Non-radiality can arise when the magnetic field in the outer corona contains, in addition to the dipole, one or more higher order harmonic components of comparable strength \citep{wang_nonradial_1996}. The current sheet bends in latitude because the different multipoles, each with its own particular neutral-line topology, decrease with the radial distance at different rates. This could also be an explanation for the difference found in the latitude for the streamer, where we locate the peak of brightness in an arc-shaped profile (see Section~\ref{s:position}), and the location of the neutral line on the source surface field map. PSFF extrapolation, where the field lines are required to be purely radial beyond the source surface, can not reproduce the non-radial streamers.

\begin{figure}
    \centering
    \epsscale{1.1}
    \plotone{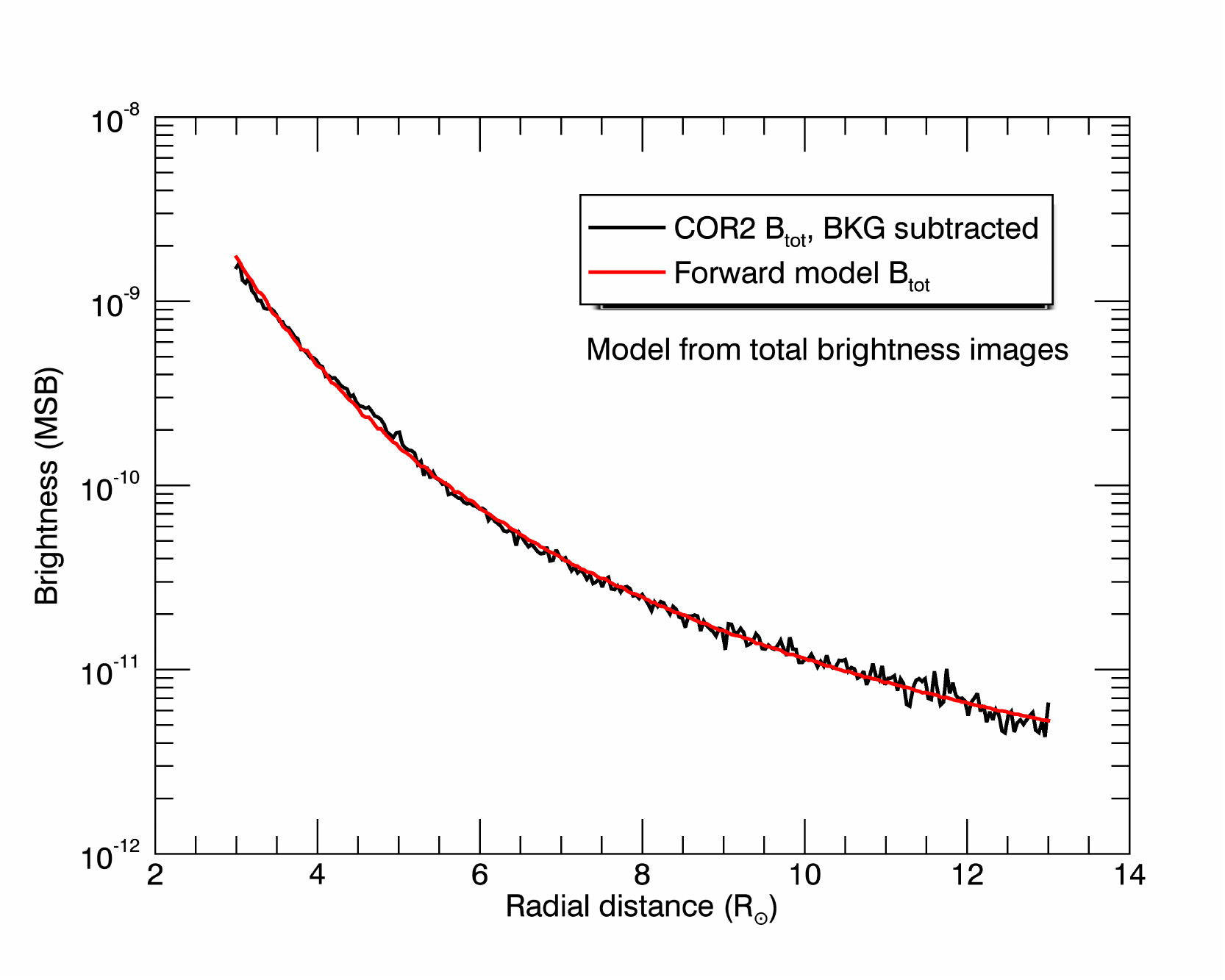}
    \caption{Radial profiles of the K corona brightness along the streamer viewed edge-on by COR2 aboard STEREO~A, derived from the total brightness, background subtracted image. The position angle of the profile is 129\degr{} in the STEREO~A/COR2 image. The black and red lines correspond to the data and the model fitted with total brightness images, respectively.}
    \label{fig:radial}
\end{figure}

Next, we continue with the independent fit in the radial direction by using one radial profile along the streamer in the edge-on view at P.A. 129\degr{} (line R in Figure~\ref{fig:extractions}) and one arc-shaped profile along the face-on view at 3~$R_{\sun}$ (curve A2 in Figure~\ref{fig:extractions}) to determine the parameters $a_i$ from the $n_{\mathrm{e, radial}}$ function. Since the range of the radial brightness profile covers a few orders of magnitude in the COR2 field of view, we modify the profile by taking the logarithm before implementing it in a $\chi^2$ criterion on which we perform a minimization. We use the SCRaytrace raytracing software available in SolarSoft \citep{2004AGUFMSH21B0404T} to create a synthetic view of our density model by integrating along the line-of-sight, corresponding to the coronagraph we used in the observation. The electron density inversion used in this software is based upon the method presented in \citet{hayes_deriving_2001}. The brightness due to Thomson-scattering relates to the electron density. This relation is inverted along radial profiles to obtain the forward models. We then obtain the reconstructed profiles by extracting the same profiles that we chose in the observations from these forward modeled views. The following criteria is then minimized:
\begin{equation}
\boldsymbol{\hat{a}} = \operatorname*{arg\,min}_{\boldsymbol{a}} \sum \left[ \log (B) - \log (\tilde{B}(\boldsymbol{a})) \right]^{2} ,
\end{equation}
where $B$ is the radial brightness profile of the edge-on image, and $\tilde{B}$ is the reconstructed radial brightness profile of the forward modeled view; $\boldsymbol{a}$ refers to the vector of the coefficients $a_i$.
Again the minimization is done with the multivariate MPFIT algorithm. The result for the fit of the radial profile can be seen in Figure~\ref{fig:radial} (solid lines). The radial profile of the modeled brightness is in good agreement with the COR2 A observations.

\begin{figure}
    \centering
    \epsscale{1.1}
    \plotone{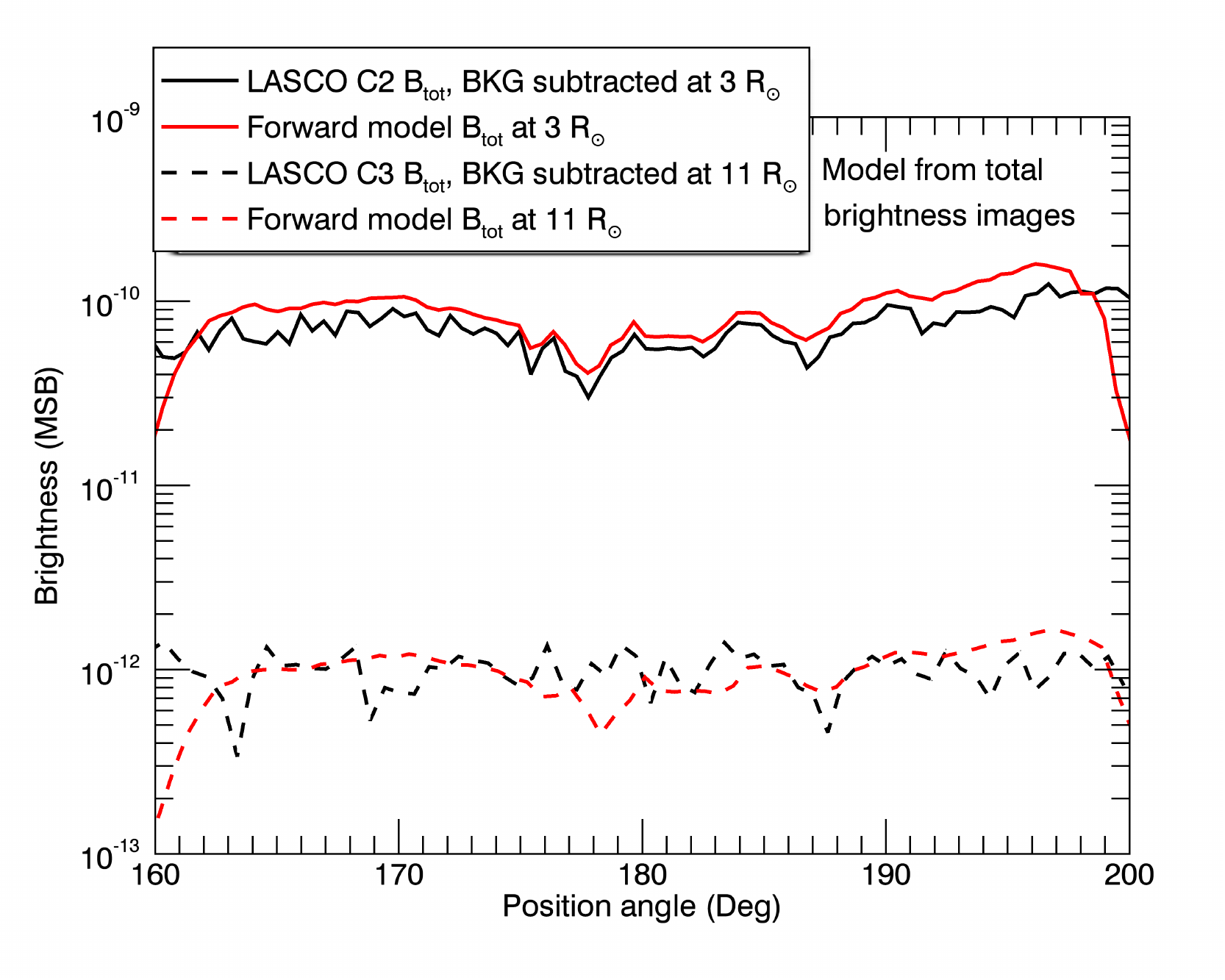}
    \caption{Face-on (azimuthal) profiles of the brightness of the streamer at 3 (solid lines) and 11~$R_{\sun}$ (dashed lines) in the face-on view (curves A2 and A3 in Figure~\ref{fig:extractions}). The black and red brightness profiles are related to the total brightness, background subtracted LASCO observations and the forward model calculated using the optimized $n_{\mathrm{e}}$ to the total brightness images, respectively.}
    \label{fig:azimuthal}
\end{figure}

We also compare the azimuthal profiles from the LASCO observations (arcs A2 and A3 in Figure~\ref{fig:extractions}) to reconstructed azimuthal brightness profiles from corresponding forward modeled views of the model. The observed brightness enhancements and depletions near the center of the streamer slab are quite nicely reproduced by our model at the height of our chosen profile, as can be seen in Figure~\ref{fig:azimuthal}. At the edges, the features seem to become smeared out to comply with the condition that the density must go to zero outside of the streamer slab. In this respect, our model behaves similarly to the model by \citet{thernisien_electron_2006}, see their Figure~10. Even though the absolute values of intensity are only incorporated in the model through the radial profile of the COR2 view, we get a very good match for the absolute values of intensity in the LASCO C2 and C3 field of view, as can be seen in the profiles extracted from the model at 3 and 11~$R_{\sun}$. At 11~$R_{\sun}$, not all the variations in the profile correspond to variations that can be seen in the input profile at 3~$R_{\sun}$, so we do not capture all the variations at heights other than the one of our chosen profile. This indicates that not all the ray-like structures in the LASCO view are perfectly radial.

\begin{deluxetable}{c  c  c  c  c  c }[b]
	\tablecaption{Summary of the fitting parameters for the set of total brightness images \label{tbl:overview}}     
	\tablehead{\colhead{} & \multicolumn{5}{c}{Subscript} \\
	\colhead{Parameter} & \colhead{0} & \colhead{1} & \colhead{2} & \colhead{3} & \colhead{4} }
	\startdata
	$a_i$ & \ldots & 1.736e5 & -2.999e4 & 5.762e6 & 9.699e7 \\
	$b_i$ & -87.79 & 1105. & -4785. & 8035. & -4967. \\
	$c_i$ & 3.927 & -3.623 & 136.4 & -1064. & 2409.
	\enddata
\end{deluxetable}

A summary of the values for the different parameters used in our optimization process can be found in Table~\ref{tbl:overview}.

\subsection{Fitting the polarized brightness images}\label{s:pbparameters}
We now repeat the process described in the previous section, but for the set of \textit{pB} images from LASCO~C2 and COR2. Again, we first fit the shape function to seventeen normalized arc-shaped profiles taken at different heights in the edge-on view. Then, we continue with the fit in the radial direction using one radial profile along the streamer in the edge-on view at P.A. 129\degr{} (line R in Figure~\ref{fig:extractions}) and one arc-shaped profile along the face-on view at 3~$R_{\sun}$ (curve A2 in Figure~\ref{fig:extractions}). For the minimization however, this time we calculate the \textit{pB} forward modeled profiles, and fit these to the observed \textit{pB} profiles. A summary of the values for the different parameters resulting from this minimization can be found in Table~\ref{tbl:overviewpb}.

\begin{figure}
    \centering
    \epsscale{1.1}
    \plotone{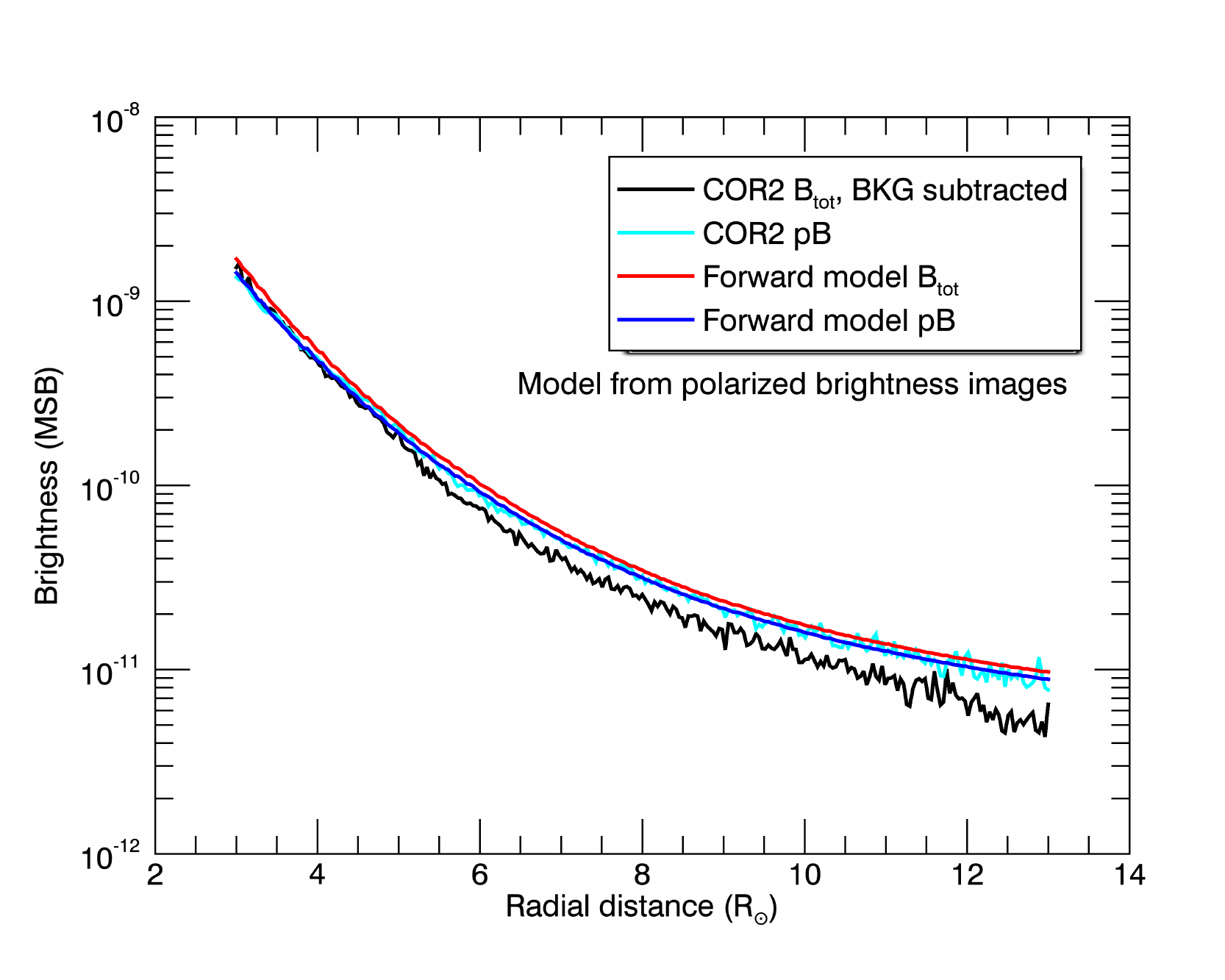}
    \caption{Radial profiles of the K corona brightness along the streamer viewed edge-on by COR2 aboard STEREO~A. Data is derived from the total brightness, background subtracted image (black) and the \textit{pB} image (cyan). The red and blue lines correspond respectively to the total and polarized brightness calculated with the model, fitted with \textit{pB} images. The position angle of the profile is 129\degr{} in the STEREO~A/COR2 image.}
    \label{fig:radial_pb}
\end{figure}

\begin{figure}
    \centering
    \epsscale{1.1}
    \plotone{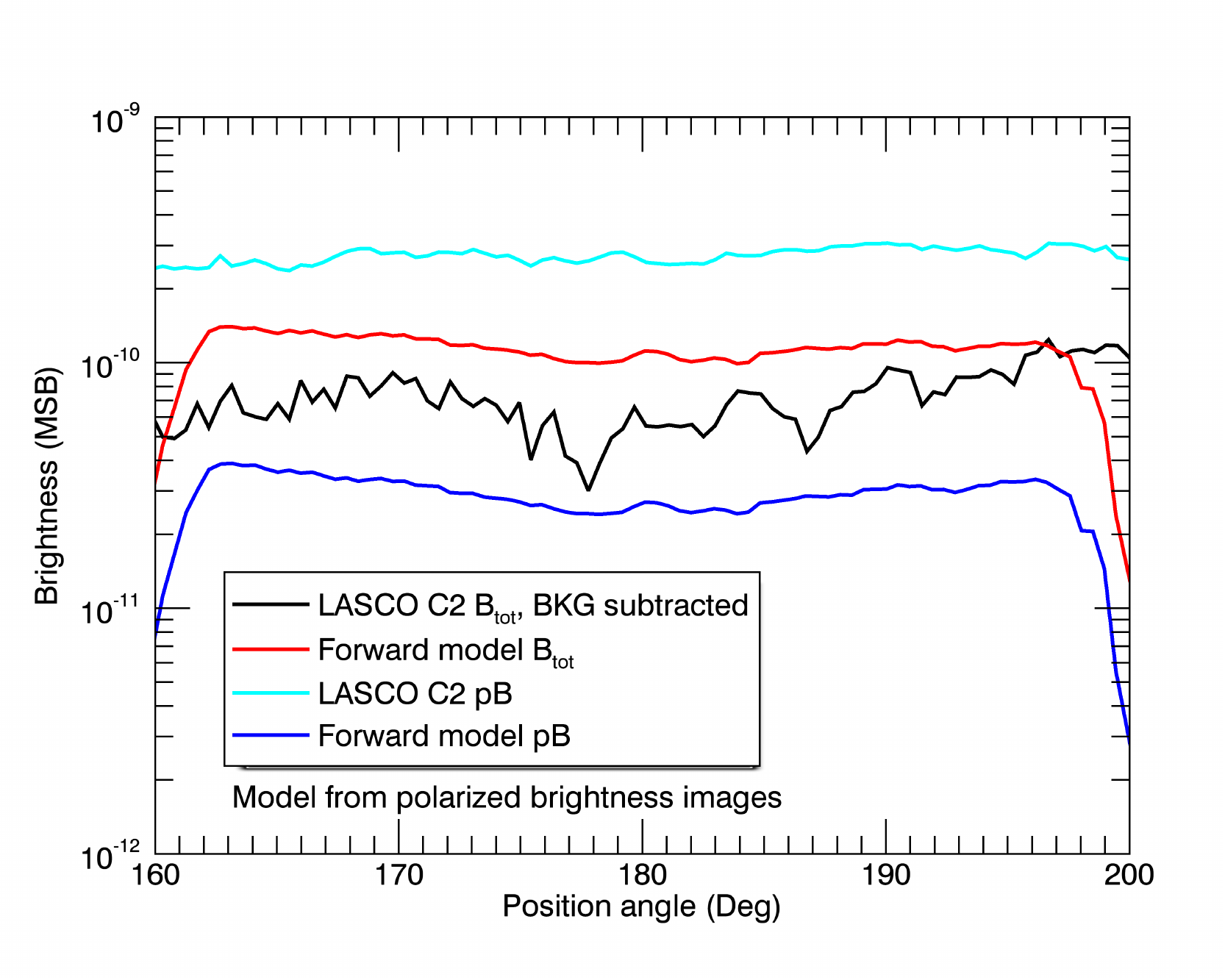}
    \caption{Face-on (azimuthal) profiles of the brightness of the streamer at 3~$R_{\sun}$ in the face-on view (curve A2 in Figure~\ref{fig:extractions}). Data is derived from the total brightness, background subtracted image (black) and the \textit{pB} image (cyan) taken by SOHO/LASCO~C2. The red and blue lines correspond respectively to the total and polarized brightness calculated with the model, fitted with \textit{pB} images.}
    \label{fig:azimuthal_pb}
\end{figure}

To compare this model to the observations, we calculated the forward modeled views of \edit1{\added{both}} the \textit{pB} \edit1{\added{and the total brightness from the model with the \textit{pB} images as input,}} and extracted the corresponding profiles. Figure~\ref{fig:radial_pb} shows the observed and forward modeled \textit{pB} profiles along the radial line R (as seen in Figure~\ref{fig:extractions}), as the cyan and blue line respectively. \edit1{\added{The red line corresponds to the total brightness profile obtained with the model from \textit{pB} images, and the black line is again the observed profile from the total brightness, background subtracted image.}} We find that our \edit1{\added{\textit{pB} profile from the}} model fits the \edit1{\added{observed \textit{pB}}} radial profiles very well. \edit1{\added{We provide the corresponding four curves for the azimuthal view (along arc A2 in Figure~\ref{fig:extractions}) in Figure~\ref{fig:azimuthal_pb}, that is the forward modeled total brightness and \textit{pB} profile from the model with \textit{pB} images as input, the observed profile from the \textit{pB} image, and the observed profile from the total brightness, background subtracted image.}} For the azimuthal profile at 3~$R_{\sun}$, Figure~\ref{fig:azimuthal_pb} shows that there is a significant difference in contrast between the small-scale structures in the \edit1{\added{observed}} total and polarized brightness profiles. For the total brightness\edit1{\added{, background subtracted}} profile, we see clear enhancements and depletions, but they are much less pronounced in the profile from the \textit{pB} observations. In general, the signal-to-noise ratio in the \textit{pB} images is lower than that in the total brightness images, which could explain why these variations are less pronounced in the \textit{pB} image.  In addition, in both Figure~\ref{fig:radial_pb} and Figure~\ref{fig:azimuthal_pb} the observed \textit{pB} curve is slightly above the total brightness curve, which means that probably too much of the K corona was removed during the background subtraction (see Section~\ref{s:calibrating}). From Figure~\ref{fig:azimuthal_pb}, it is also clear that the fit of the forward modeled \textit{pB} profile is significantly lower than the observed \textit{pB} profile in LASCO C2, although the modeled total brightness curve roughly matches the background subtracted data. \edit1{\deleted{We will discuss this discrepancy in the next section.}}

\begin{deluxetable}{c  c  c  c  c  c }[b]
	\tablecaption{Summary of the fitting parameters for the set of \textit{pB} images \label{tbl:overviewpb}}     
	\tablehead{\colhead{} & \multicolumn{5}{c}{Subscript} \\
	\colhead{Parameter} & \colhead{0} & \colhead{1} & \colhead{2} & \colhead{3} & \colhead{4} }
	\startdata
	$a_i$ & \ldots & 6.866e5 & -1.047e7 & 7.557e7 & -7.696e7 \\
	$b_i$ & -73.61 & 841.8 & -2710. & 1956. & 1827. \\
	$c_i$ & -1.120 & 128.9 & -1105. & 3473. & -3302.
	\enddata
\end{deluxetable}


\begin{figure}
    \centering
    \epsscale{1.1}
    \plotone{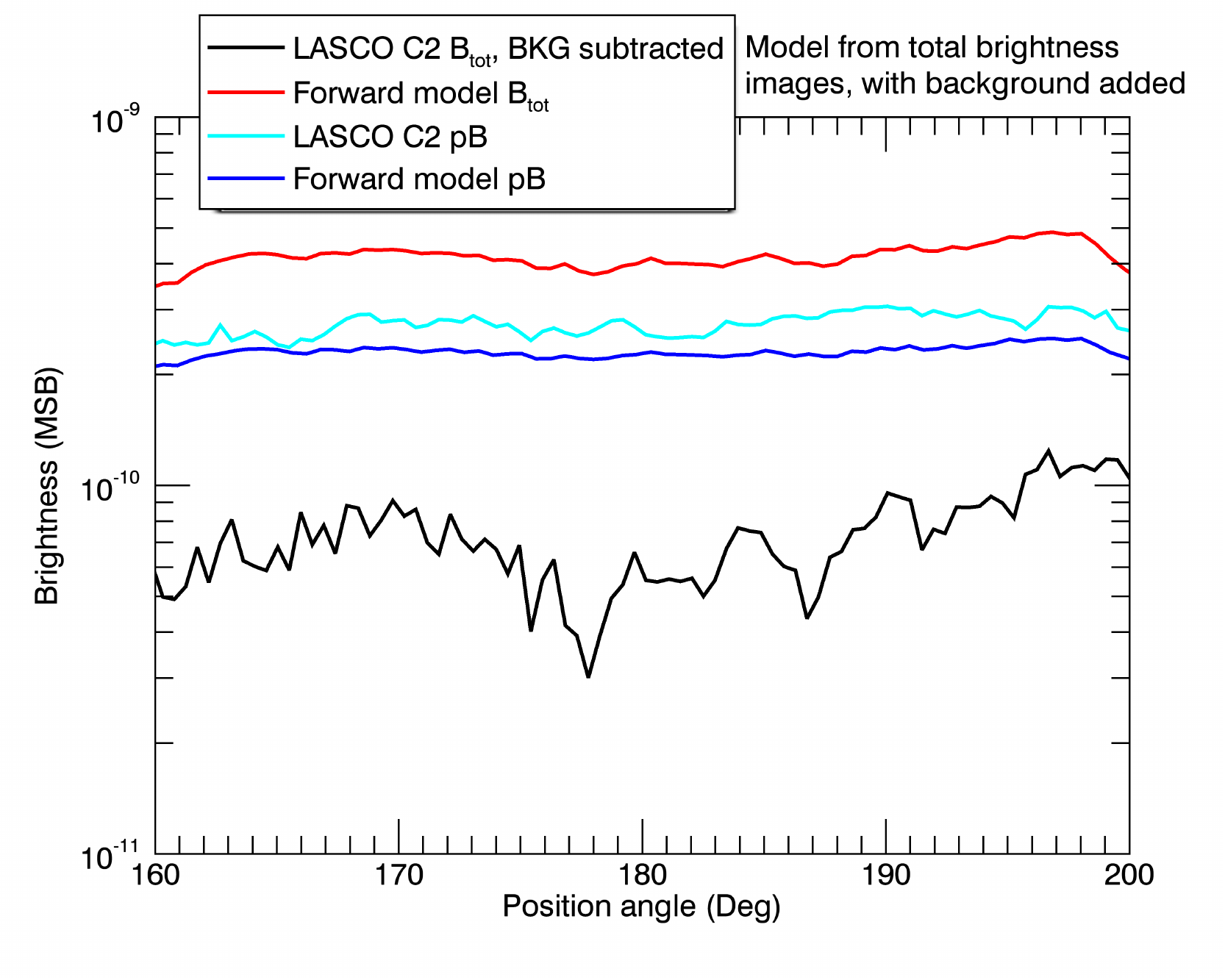}
    \caption{Azimuthal profiles of the streamer brightness at 3~$R_{\sun}$ in the face-on view (curve A2 in Figure~\ref{fig:extractions}). Data is derived from the SOHO/LASCO~C2 total brightness, background subtracted image (black) and the \textit{pB} image (cyan). These curves are the same as in Figure~\ref{fig:azimuthal_pb}. The red and blue lines correspond respectively to the total and polarized brightness calculated with the model, to which a polar density from \citet{guhathakurta_large-scale_1996} is added and total brightness images are fitted.}
    \label{fig:azimuthal_g}
\end{figure}

\edit1{\added{This discrepancy could be partly explained by the high sensitivity of LASCO C2 \textit{pB} observations to small changes in the calibration factors. \cite{morgan2015} argued that the calibration factors should be modified slightly in comparison to the standard SolarSoft calibration factors that we used in the present study. This could lead to a noticeable decrease of the observed \textit{pB}, especially above the poles. The calibration factor modification would lead to a decrease of the observed \textit{pB} in Figure~\ref{fig:azimuthal_pb} since the arc A2 for which we extract the azimuthal profiles is situated above the south pole in the field of view of LASCO C2.}}

\edit1{\deleted{Since the results of our two models are consistent with each other and with previous results, we believe that the origin of a poor correspondence of the observed and modeled \textit{pB} profiles in Figure~\ref{fig:azimuthal_pb} must be searched elsewhere. }\replaced{A}{Another} plausible explanation \added{for this discrepancy} is that there is a static K corona component, comparable to the intensity values in the LASCO field of view, which contributes to the \textit{pB} images in the LASCO field of view. If this additional component is located around the south pole (i.e. close to the plane of the sky in LASCO images), then it would provide a significant contribution to the LASCO C2 image. However, this component would not be significant in the COR2 field of view, since it would be located out of the plane of the sky at the lines of sight passing through the streamer, and the intensity of the streamer is much larger there. This static component is a real part of the K corona, but it gets subtracted with our background removal procedure. To test this hypothesis, we added an additional global density to our model from the total brightness images, using the model by \citet{guhathakurta_large-scale_1996}. In their equation (10), we replaced the current sheet density term $N_{\rm cs}(r)$ with our streamer model and implemented the polar density term $N_{\rm p}(r)$ as listed in their Table 3. The calculations using this combined streamer+pole density model show that the results for the COR2 field of view have almost not changed and the resulting $pB$ curve is very close to the blue curve in Figure~\ref{fig:radial_pb}. However, in the LASCO field of view, the calculated \textit{pB} has increased significantly and now matches the observed values much more closely, as can be seen in Figure~\ref{fig:azimuthal_g}. We can also see that this added density in the polar region smoothens the variations present in the azimuthal profile. This static density component could thus also explain the lower contrast of the ray-like structures in the \textit{pB} observations. }

\edit1{
This combined model is used here only for illustration as we aim only at constructing a local density model for the streamer, and not creating a global coronal electron density model to fit the full observed image. For such a global model, one would need to fit also the static component of the K corona. \edit1{\replaced{However, it is impossible to separate correctly the K corona}{Separating the}} static components \edit1{\added{of the K corona}} and the ray-like structures in the streamer \edit1{\added{is difficult}} with the currently available data \edit1{\added{and would require a more advanced model, which goes beyond the scope of the present work. }}}

\section{Discussion}\label{s:discussion}

\subsection{Comparison of fitting with total and polarized brightness images}\label{s:comparisontwomodels}
With the profiles shown in Figure~\ref{fig:radial_pb} and Figure~\ref{fig:azimuthal_pb}, we can first compare the total brightness observations with the polarized brightness observations. Below 5~$R_{\sun}$ in the COR2 field of view, we see that the observed radial \textit{pB} profile (cyan line in Figure~\ref{fig:radial_pb}) falls slightly above the observed total brightness profile. As noted before, this is probably due to subtracting too much of the K corona. Above this height, we know that the F corona becomes polarized, and therefore starts adding to the \textit{pB}. Since we still subtract the static F corona in the total brightness image through the background subtraction, the total brightness profile significantly falls below the \textit{pB} profile. From the total brightness and \textit{pB} profiles generated by the model (red and blue curves in Figure~\ref{fig:radial_pb}), we can see that in this particular streamer configuration, it is expected that the total brightness from only the K corona lies very close to the \textit{pB} and thus that all modeled and observed profiles fall very close to each other below 5~$R_{\sun}$. The profiles at 3~$R_{\sun}$ in the LASCO C2 face-on view tell a different story, as shown in Figure~\ref{fig:azimuthal_pb}. The observed profile from the total brightness image with the background subtracted (black) is much lower than the observed profile from the \textit{pB} image (cyan). This is contrary to what one would expect, since polarized brightness should be always lower than the total brightness. This can be seen from the forward modeled total and polarized brightness profiles (red and blue curves in Figure~\ref{fig:azimuthal_pb}). We also notice that the polarization degree in the LASCO view is lower than that in the COR2 view. This is explained by the orientation of our streamer. \edit1{\added{In the field of view of COR2, the streamer density is strongly concentrated close to the plane of the sky which causes the Thomson-scattered emission to be more polarized.}} In the LASCO view \edit1{\added{however}}, the density is mostly concentrated out of the plane of the sky, and thus is less polarized. \edit1{\added{As the distance from the plane of the sky increases, the polarized brightness decreases faster than the total brightness \citep{deforest2013}, which explains the difference in the polarization degree in COR2 and LASCO views of the modeled streamer. }}

\begin{figure*}
	\centering
	\epsscale{0.7}
	\plotone{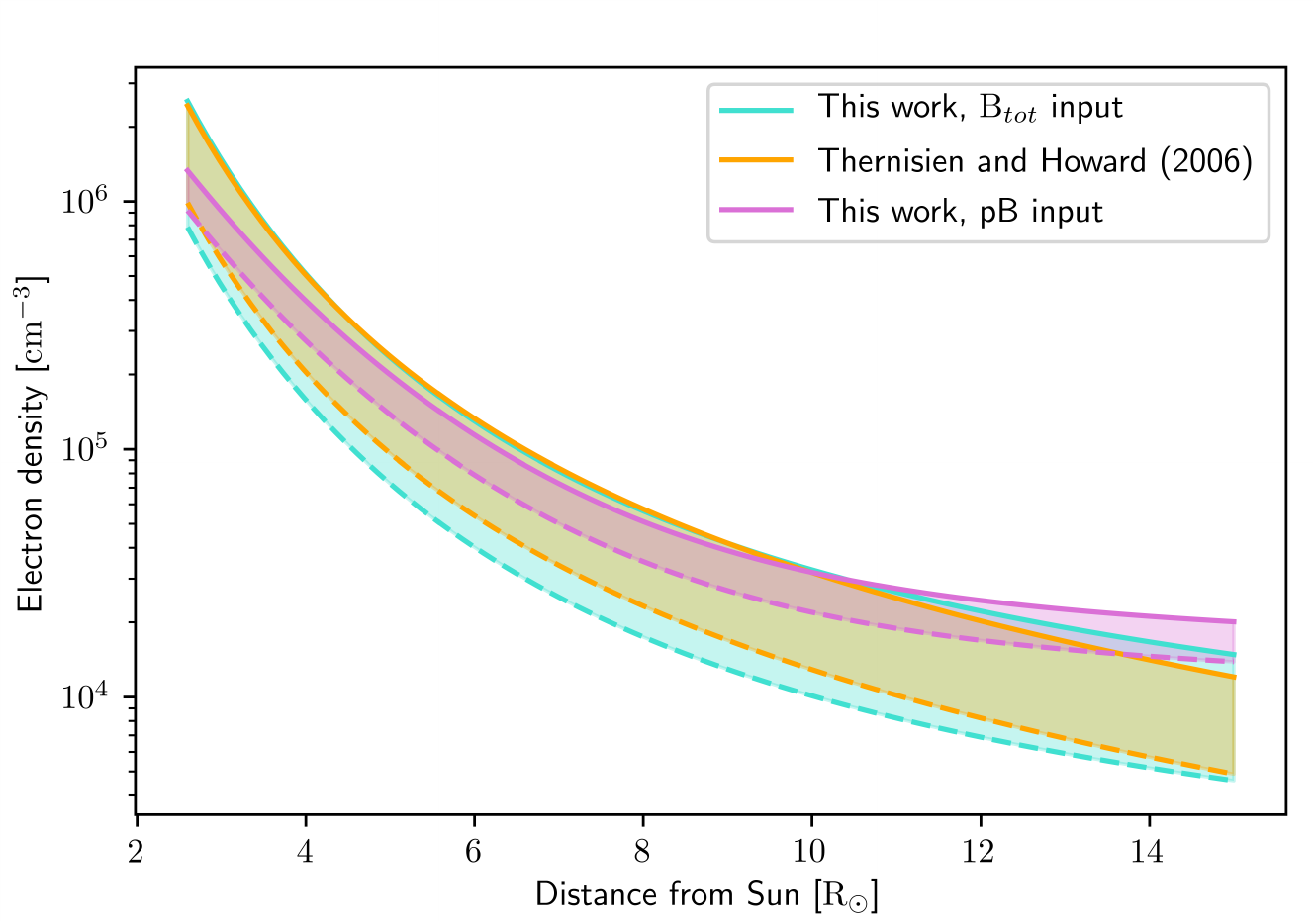}
	\caption{Coronal electron density profiles along the radial direction derived using our streamer model from total brightness and \textit{pB} images (blue and magenta, respectively) compared to the values obtained by \citet{thernisien_electron_2006} (orange). The solid (dashed) lines correspond to the maximum (minimum) density values in the plasma slab. The shaded zone presents the range of the density profiles in the bright ray-like structures visible in the face-on view.}\label{fig:electron_density}
\end{figure*}
\begin{figure}
	\centering
	\plotone{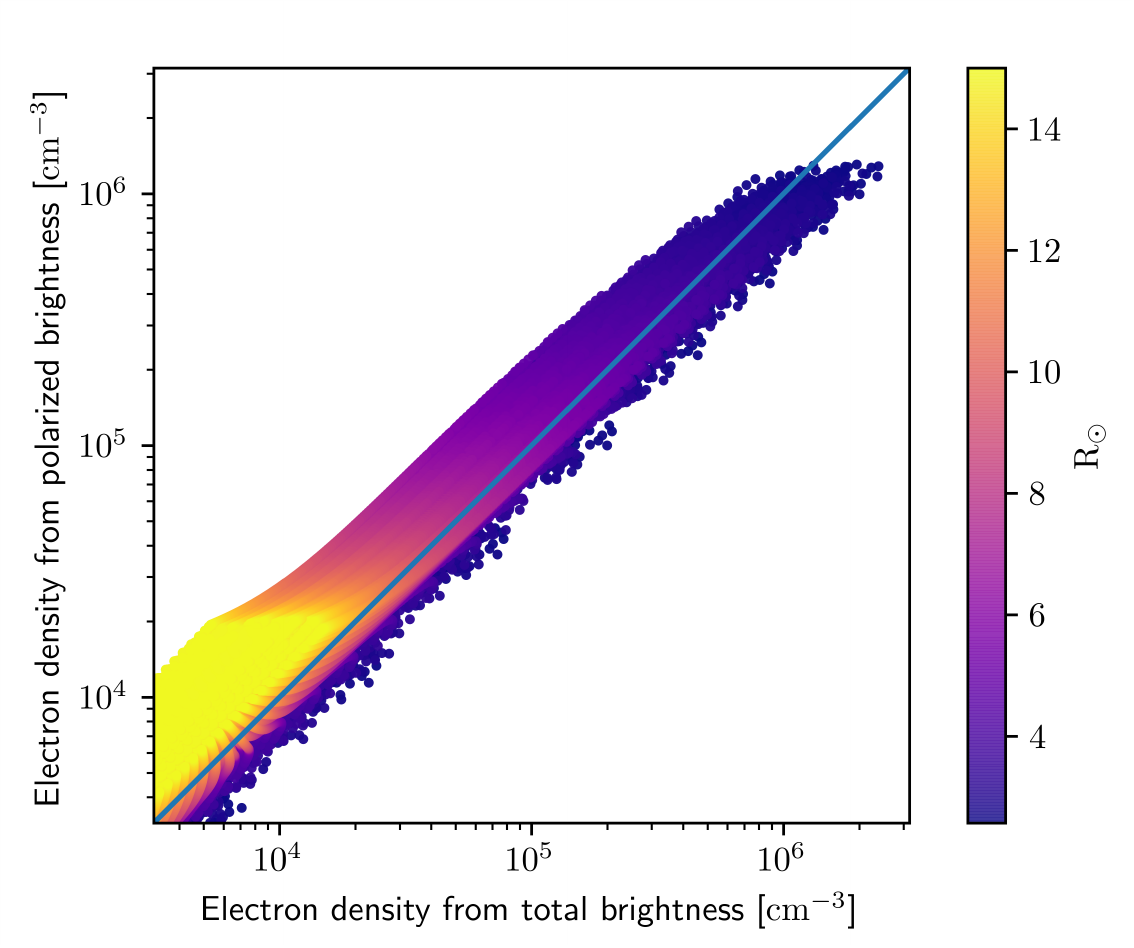}
	\caption{Scatterplot of the electron density resulting from two models: fitting to total brightness images versus fitting to \textit{pB} images. Points are colored according to their distance from the solar surface. A line illustrating the same values in both models is shown in blue for reference.}\label{fig:scatterplot}
\end{figure}

Next, we compare the densities derived from both fittings, to see how well the two models correspond to each other. Figure~\ref{fig:electron_density} shows the comparison between the radial profiles of our electron density derived from the total brightness (blue) and \textit{pB} (magenta) images, and those obtained by \citet{thernisien_electron_2006} (orange). The solid and dashed lines show the maximum and minimum electron density profiles of each modeled streamer, respectively. The electron density profiles range between the maximum and minimum values, as is highlighted by the shaded zones. For the model derived from the total brightness images, this presents a maximal density contrast of bright ray-like structures around a factor of 3 with respect to the background streamer density. Since the brightness variation in the face-on profile in the \textit{pB} images is lower, the maximal density contrast for the model from the \textit{pB} images is also lower, at about 1.5. Below 5~$R_{\sun}$,  there is a very good correspondence between the two density models. Above 5~$R_{\sun}$, we see that the model from \textit{pB} images gives higher densities than the one from the total brightness images, which is to be expected from the observations. In Figure~\ref{fig:scatterplot}, we show a scatterplot of the two density cubes, where each point is colored according to its distance from the solar surface. This plot shows that there is a very high correlation between the density cubes derived from the two models, which is also indicated by the correlation coefficient of 0.96. We also see that the correspondence between the two density cubes gets better the closer one approaches the solar surface. Farther out from the Sun, the density values for the density cube from the \textit{pB} images are too high in comparison with those from the density cube from the total brightness images. This can again be attributed to the F corona becoming polarized higher up in the corona, and thus adding to the density model from the \textit{pB} images. We can conclude that our two models are very consistent with each other regarding the resulting values for the electron density.  

Figure~\ref{fig:electron_density} shows that our results are also consistent with the density values found previously by \citet{thernisien_electron_2006}. In their paper, a comparison was also made between their model and previous density models \citep{saito_study_1977, leblanc_tracing_1998, quemerais_two-dimensional_2002}, demonstrating a good agreement with them. \edit1{\added{Since the results of our two models are consistent with each other and with previous results, we believe that the origin of a poor correspondence of the observed and modeled \textit{pB} profiles in Figure~\ref{fig:azimuthal_pb} must be searched elsewhere, as was explained in the Section~\ref{s:pbparameters}.}}

\subsection{3D rendering of the density and forward modeled views}
\begin{figure}
	\centering
	\epsscale{1.1}
	\plotone{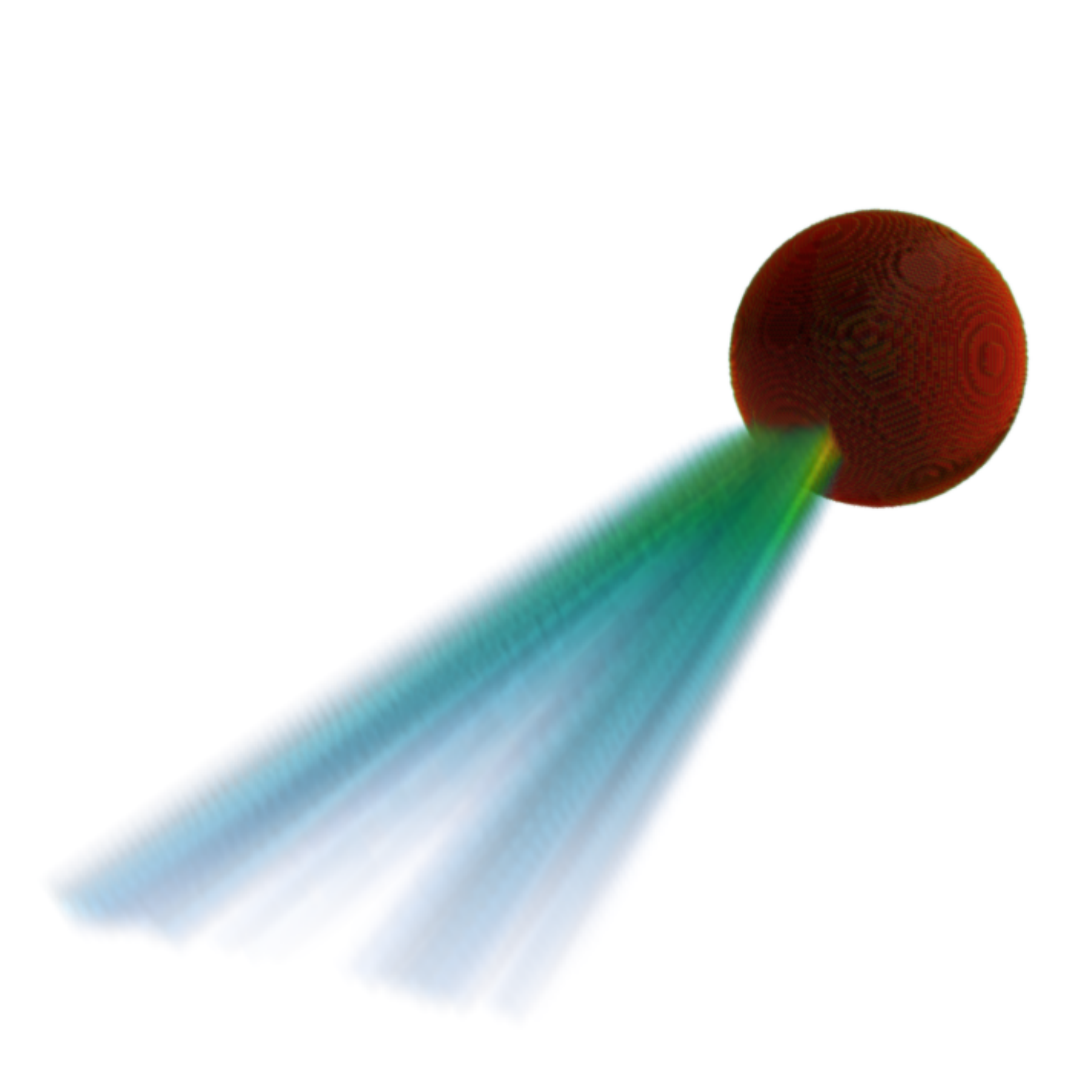}
	\caption{Three-dimensional rendering of the density cube resulting from our parameter fits to the streamer model. The density cube ranges from -15 to 15~$R_{\sun}$ in all three directions, which corresponds to the FOV of COR2. In each direction we have 512 pixels, which is half of the resolution that was used for the COR2 observations that serve as input to the model (a lower resolution was chosen to make the rendering reasonably faster). The coloring gives a representation of the scaling of the density in a logarithmic scale, but was capped to enhance visibility of different ray-like features in the plane of the slab. The orb in the center surrounds the Sun and has a radius of 2.5~$R_{\sun}$.}\label{fig:3D}
\end{figure}

Figure \ref{fig:3D} presents the slab in an isometric view of the 3D density cube obtained from the fitting to the total brightness images. This Figure gives a better understanding of how the slab looks in three dimensions. The edge-on view is actually a superposition of all the different ray-like features that form the slab structure. The rays have the rate of radial brightness decay that is similar to the rate of radial decay that can be seen in the edge-on view.
Figure \ref{fig:3D} also clearly illustrates our assumption, namely that all the rays visible in the face-on view are situated inside the streamer slab. In reality, some of the radial features seen in the face-on view may be polar plumes, or the quasi-radial density enhancement that we can see slightly southward of the streamer in the edge-on view (Figure~\ref{fig:views}, left panel), or other structures in the solar corona.

\begin{figure*}
    \centering
   	\plotone{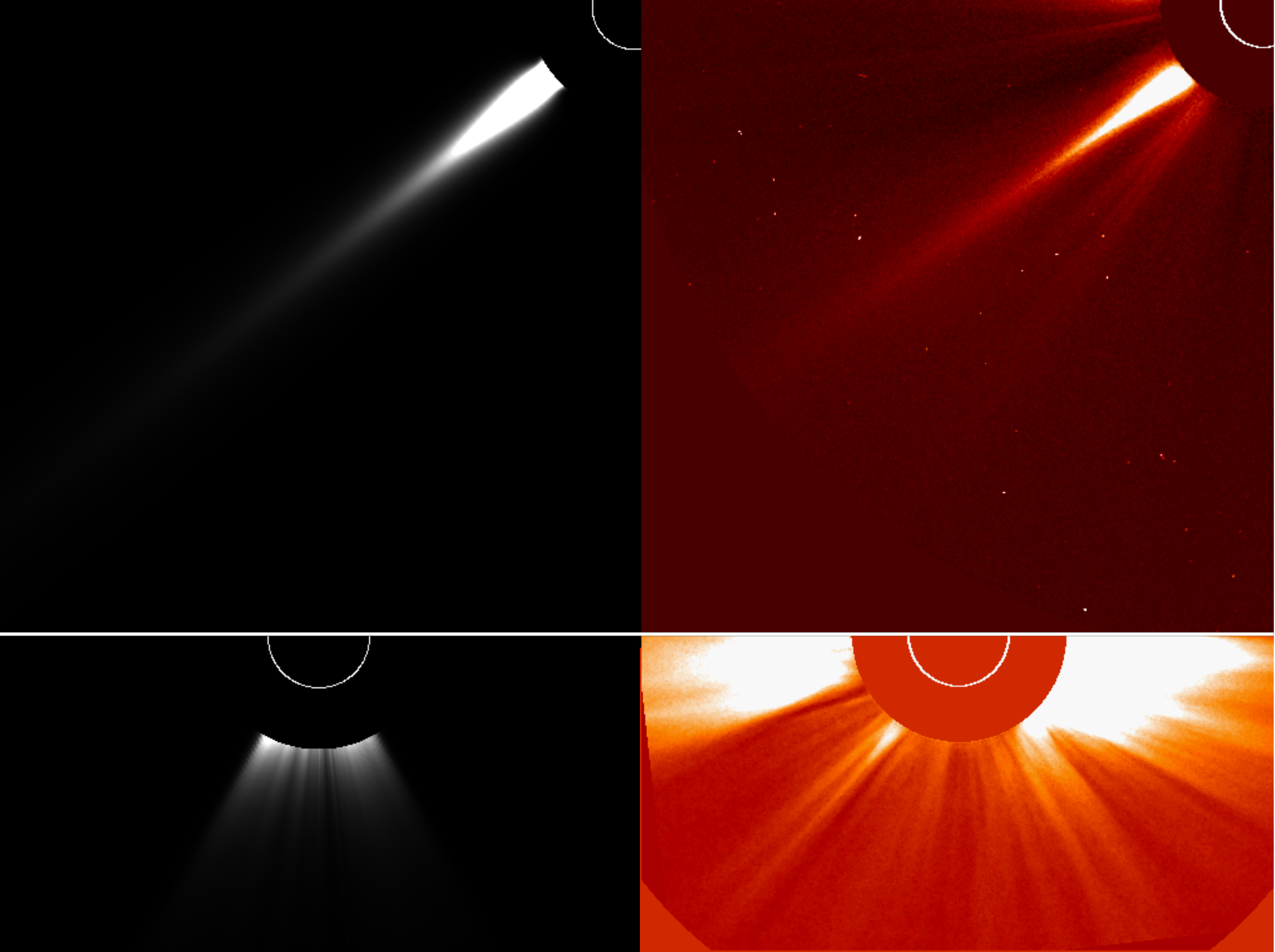}
    \caption{Forward modeled views of the obtained 3D density cube as they would appear observed by COR2 A (top) and LASCO C2 (bottom) coronagraphs (left), compared to the corresponding observations (right).}
    \label{fig:comparison}
\end{figure*}

In Figure~\ref{fig:comparison} we show the forward modeled views of the streamer from which we obtained the reconstructed profiles for the fitting to the total brightness images. They present the density cube as it would be seen by a coronagraph from two vantage points in quadrature. The central bright feature in the edge-on view resembles the observed streamer rather well. The main difference between the model and the observations is the slight non-radiality of the streamer. The face-on views are harder to compare, due to low signal-to-noise ratio of the observations. Some common features can clearly be distinguished. The bright structure to the south of our streamer in the edge-on view probably corresponds to the bright structure just outside of our slab on the right in the face-on view. The bright structures to the right of our slab in the LASCO view are too bright to lie in the plane of the slab, and must be closer to the plane of the sky. From the COR2 synoptic map (Figure~\ref{fig:synopticmap}), we can also see that there is a second streamer structure crossing the plane of the sky before our streamer (situated around Carrington longitudes 150-180\degr{}), which probably corresponds to this feature. This streamer is significantly out of the plane of the sky on the day of our observations (April 30, 2011; see Figure~\ref{fig:views}).

In Section~\ref{s:position}, we located the streamer slab following the neutral line in the PFSS extrapolation map. The high variability we now have in our density model with a number of ray-like structures along the slab, indicates that the plasma sheet is not a smooth layer centered at the current sheet. This may have implications for the magnetic field structure inside the streamer, which needs to be explained theoretically. \citet{wang2} proposed that the streamer plasma sheet is filled with coronal material due to interchange reconnection with closed magnetic loops at the streamer cusp. If this reconnection is not uniform in space and time, different parts of the plasma sheet may be filled with plasma at different times, producing the pattern with ray-like structures similar to that reported in our work. The ray-like structure would then be intrinsically dynamic. The interchange reconnection at the streamer cusp was modeled by \citet{higginson}, but considering only the magnetic (not density) structure. 

It is difficult to confirm the occurrence of this process observationally with the current instrumentation. There is a significant gap between the fields of view of externally occulted coronagraphs (like COR2) and EUV imagers (like Extreme UltraViolet Imager aboard STEREO) that is covered only by internally occulted coronagraphs (like STEREO/COR1), which are prone to high straylight and do not allow observations of fine coronal structures at sufficient resolution. We checked the data taken by the MLSO/Mk4 coronagraph \citep[that has the field of view connecting those of LASCO C2 and disk EUV imagers, see e.g.][]{mk4} but could not identify any structures corresponding to rays we observed in the C2 field of view.  Future missions like the ASPIICS coronagraph aboard PROBA-3 \citep{proba32, proba33, proba3} will fill this observational gap and have a potential to improve our knowledge of how these ray-like structures relate to the typical cusp structure that helmet streamers have in the low corona. 

\section{Conclusions}\label{s:conclusions}

We have analyzed the 3D structure of a coronal streamer using a forward model based on plausible assumptions about the large-scale 3D geometry. The streamer is represented by a dense plasma slab with a radial density decrease and azimuthal fine structure. We fitted this model to both total and polarized brightness data of a streamer, observed by STEREO~A/COR2 and SOHO/LASCO C2 and C3 coronagraphs, while the STEREO and SOHO spacecraft were in quadrature. Our model can reproduce the observations reasonably well, as shown by the forward model calculations of the streamer images in the two different views and the fits to the observed brightness profiles. The two sets of fittings (one to the total brightness background-subtracted images, and one to the \textit{pB} images) are consistent with each other, and we also obtained a good agreement with earlier streamer density models. The assumption that coronal streamers are purely radial features is not fully correct, which could be improved by expanding the model to allow for non-radial streamers. 

The model sheds light on the fine three-dimensional structure of the electron density distribution in coronal streamers. The ray-like structures in the slab model are necessary to reproduce the observations in the face-on view. We found variations up to a factor of 3 between the radial profiles of the electron density of brighter and darker structures in the streamer.

Using total brightness images, from which we subtracted a minimum background, is not ideal for separating the K and F corona. We have shown that below approximately 5~$R_{\sun}$, the model obtained from the total brightness images corresponds very well to the model obtained from the \textit{pB} images. Although the subtraction of a minimum image is not an ideal method for separating the K and F corona, we demonstrate it to be a decent method to obtain density values inside coronal streamers. The derived densities are within a factor 3 from the densities obtained from the model using \textit{pB} data as input. Above 5~$R_{\sun}$ however, polarized brightness data is no longer suitable due to the polarization of the F corona. A better understanding of the F corona in this region would certainly improve the K and F corona separation, and refine methods for the inversion of total and polarized brightness images into electron density.

The density model developed here can be used as a background density model for modeling of e.g. streamer wave events. Nevertheless, the method should be further expanded to cases where different available coronagraphs are not in quadrature to be able to use it for specific events found during the epoch of STEREO observations. \edit1{\added{Further development of the model could go towards combining it with more global coronal models (e.g. those derived from tomography). At the end of Section~\ref{s:pbparameters}, we gave an illustration of how the model better describes the observations when it is implemented in a simple global coronal density model. This could be explored more by optimizing the static density component to fit the observations, and by using more advanced models for the static density component. This way, our model could locally provide finer details in a global coronal density model. An advanced combined model could then shed light on the role of the fine structure of streamers in the global corona, and how it relates to the slow solar wind formation.}}

%
\acknowledgments
The authors thank the anonymous referee for the valuable comments which led to significant improvements of this paper. ANZ thanks the European Space Agency (ESA) and the Belgian Federal Science Policy Office (BELSPO) for their support in the framework of the PRODEX Programme. TVD was supported by the GOA-2015-014 (KU~Leuven) and the European Research Council (ERC) under the European Union's Horizon 2020 research and innovation programme (grant agreement No 724326). We thank A. Thernisien, M. Mierla and R. Colaninno for useful discussions. The SECCHI data used here were produced by an international consortium of the Naval Research Laboratory (USA), Lockheed Martin Solar and Astrophysics Lab (USA), NASA Goddard Space Flight Center (USA), Rutherford Appleton Laboratory (UK), University of Birmingham (UK), Max-Planck-Institut for Solar System Research (Germany), Centre Spatiale de Li\`ege (Belgium), Institut d'Optique Th\'eorique et Appliqu\'ee (France), Institut d'Astrophysique Spatiale (France). The SOHO/LASCO data used here are produced by a consortium of the Naval Research Laboratory (USA), Max-Planck-Institut for Sonnensystemforschung (Germany), Laboratoire d'Astrophysique de Marseille (France), and the University of Birmingham (UK). SOHO is a project of international cooperation between ESA and NASA. Wilcox Solar Observatory data used in this study were obtained via the web site \url{http://wso.stanford.edu} at 2018/11/16 07:41:31 PST,  courtesy of J.T. Hoeksema. The Wilcox Solar Observatory is currently supported by NASA.

%
%
\bibliographystyle{aasjournal} 
\bibliography{references}

%

\listofchanges

\end{document}